\documentclass[aps,prb,superscriptaddress,twocolumn,floatfix,showpacs]{revtex4}
\usepackage{graphicx}
\usepackage{setspace}

\begin{document}

\def\ie{{\it i.e.}}
\def\eg{{\it e.g.}}
\def\etal{{\it et al.}}

\title{Understanding the role of ions and water molecules in the NaCl dissolution process}
\author{Ji\v{r}\'\i\ Klime\v{s}}
\affiliation{London Centre for Nanotechnology and Department of Chemistry, University College London, London, WC1E 6BT, UK}
\author{David R. Bowler}
\affiliation{London Centre for Nanotechnology and Department of Physics and Astronomy, University College London, London, WC1E 6BT, UK}
\author{Angelos Michaelides}
\email{angelos.michaelides@ucl.ac.uk}
\affiliation{London Centre for Nanotechnology and Department of Chemistry, University College London, London, WC1E 6BT, UK}

\pacs{
68.43.Bc 
68.43.Hn 
}

\date{\today}

\begin{abstract}

The dissolution of NaCl in water is one of the most common everyday processes, yet it remains
poorly understood at the molecular level.
Here we report the results of an extensive density functional theory study in which the initial
stages of NaCl dissolution have been examined at low water coverages.
Our specific approach is to study how the energetic cost of moving an ion or a pair of ions to a less 
coordinated site at the surface of various NaCl crystals varies with the number of water molecules 
adsorbed on the surface.
This ``microsolvation" approach allows us to study the dependence of the defect energies on the 
number of water molecules in the cluster and thus to establish when and where dissolution becomes 
favorable.
Moreover, this approach allows us to understand the roles of the individual ions and water molecules 
in the dissolution process.
Consistent with previous work we identify a clear preference for dissolution of Cl ions over Na ions.
However, the detailed information obtained here leads to the conclusion that the process is governed by the higher affinity
of the water molecules to Na ions than to Cl ions.
The Cl ions are released first as this exposes more Na ions at the surface creating
favorable adsorption sites for water.
We discuss how this mechanism is likely to be effective for other alkali halides.

\vskip 1cm
Copyright (2013) American Institute of Physics. This article may be
downloaded for personal use only. Any other use requires prior permission of the author
and the American Institute of Physics.
The following article appeared in J. Chem. Phys. {\bf 139}, 234702 (2013) and may be found at
\href{http://scitation.aip.org/content/aip/journal/jcp/139/23/10.1063/1.4840675}{http://scitation.aip.org/content/aip/journal/jcp/139/23/10.1063/1.4840675}.

\end{abstract}

\maketitle

\section{Introduction}

Solid-liquid interfaces play a crucial role in many natural and industrial processes. 
Heterogeneous catalysis, electrochemistry, or weathering of rocks are just some of the
many examples. 
Of all the various processes at wet surfaces, the dissolution of sodium chloride 
(rock salt, NaCl) in water is one which is frequently encountered.
Although seemingly simple, gaining understanding of NaCl dissolution is becoming increasingly important because of its
relevance to processes in the environment and water purification.
It is also likely to be relevant for other water-solid interfaces and dissolution processes and thus by understanding the
mechanism of NaCl dissolution we may start to get general insight relevant to a range of related processes.

The dissolution process has been the subject of a number of experimental studies
using a wide range of techniques.
Infrared spectroscopy measurements\cite{peters1997,peters1999,foster2000} have shown that the thickness of 
adsorbed water films depends on relative humidity and 
when the coverage reaches about 3 monolayers at ambient temperature a spectrum resembling 
brine is observed suggesting the onset of dissolution of the surface. 
These studies and previous ultraviolet photoelectron spectroscopy experiments\cite{folsch1991}
stressed the important role defects play in the dissolution process.
However, as information averaged over the whole surface of the sample is obtained,
a detailed molecular scale picture is missing from this work.
Much more detailed information has been obtained using
atomic force microscopy (AFM), scanning force microscopy (SFM), or scanning polarization
force microscopy (SPFM)\cite{hu1995spfm1,hu1995spfm2} techniques in experiments  
of Dai~{\it at al.},\cite{dai1997} Shindo~{\it et al.},\cite{shindo1996,shindo1997,shindo1998}
and others.\cite{luna1998,garcia2004,ghosal2005,verdaguer2005,verdaguer2008}
The measurements revealed slow step movements and rounding of sharp kinks at humidities below $\sim$40~\%
and a faster step movement when the humidity is increased above $\sim$40~\%.
The SPFM experiment also observed that before a water film covers the terraces, 
water molecules adsorb at defects\cite{dai1997,luna1998,ghosal2005,verdaguer2005} such as steps.
At humidities of $\sim$35\% this leads to an increase in SPFM contrast which was explained by the release of ions  
from the step and their solvation. 
An observed decrease of the contact potential has been explained by assuming that the Cl ions are released first from the steps.
Similar conclusions were presented based on experiments combining AFM and ambient pressure X-ray 
photoelectron spectroscopy by Verdaguer~{\it et al.}.\cite{verdaguer2008}
However, it is not clear why the Cl should be released first since when cations are present
as impurities at the steps, they are solvated first.\cite{verdaguer2005}

As molecular level information is often not available from experiment,
molecular simulations have been instrumental in deepening the understanding of the dissolution process. 
So far, mainly two rather distinct approaches have been used:
the first uses accurate post-Hartree-Fock (post-HF) methods to calculate binding energies
of small water clusters containing Na and Cl ions. 
The use of post-HF approaches limits the size of the system and also 
it implies that zero temperature potential energy surfaces are computed.
Using this approach adsorption of water molecules 
on a small cluster was studied and the energy needed to displace one of the ions from the
ideal lattice position calculated (see, {\eg}, Ref.~\onlinecite{yamabe2000} or \onlinecite{barnett1996}). 
In related work\cite{jungwirth2000} the hydration of a single NaCl molecule and the 
change from contact ion pair to solvent separated ion pair was studied.
Despite the generally higher accuracy offered by the post-HF methods, the rather small size 
of the systems that can be treated means that the results and trends obtained might not be 
transferable to actual surfaces of NaCl crystals.

The second main approach used to study the salt dissolution process involves molecular dynamics (MD);
usually with empirical force fields (FF).  
This allows for larger systems to be examined but with lower accuracy compared to the post-HF methods.
From the various studies performed,\cite{ohtaki1988,yang2005,bahadur2006,zasetsky2008}
Yang~{\it et al.}\cite{yang2005} has recently found that for an NaCl nanocube immersed in water the Cl ions leave
the crystal first. 
Although during these simulations the dissolution process was directly addressed, the simple force fields employed 
don't account for effects such as polarization or possible dissociation of the water molecules.
Only recently have studies appeared where density functional theory (DFT) was used to study the interaction 
of liquid water films with a flat salt surface\cite{liu2008,liu2009} 
and the dissolution process into a liquid water film.\cite{liu2011}
In the latter dissolution study it was found that the Cl ion has a lower free energy barrier to dissolution and an intermediate 
partially solvated state was identified. 
However, the huge cost of an {\it ab initio} MD study at a solid-liquid interface meant that dissolution
at just one type of surface defect could be explored and with a full liquid overlayer many
of the details of the dissolution process were smeared out.

In this work we compare energies of static low energy water clusters on NaCl surfaces;
either pristine surfaces or surfaces with ions or ion pairs displaced to a low coordination site 
have been examined.
We call the corresponding energy difference a displacement energy and study its dependence on the type of displaced ion(s),
number of water molecules and the type of surface.
We take a small number of water molecules (below 20, corresponding to coverages less than half a monolayer) 
to understand the role they play and we also use realistic models of the NaCl surface.
The unprecedented number of structures allows us to understand the basic forces governing the process and ask 
questions like how many water molecules are required for the release of an ion to become energetically favorable 
and which of the ions is likely to be released first and why.
Not unexpectedly,\cite{stockelmann1999,garcia2004} we find that dissolution is not favorable from a flat surface.
However, we can confirm the release of Cl ions from steps as suggested in previous experiments and simulations.\cite{verdaguer2005}
Indeed, we find that simply on energetic grounds it becomes favorable to release Cl ions from 
steps once there are enough water molecules available to solvate them.
In stark contrast to this, the Na ions are unlikely to be released as it always costs energy to do so.
Given the stronger interaction of water molecules with the Na ion, these findings are somewhat unexpected.
However, we explain this by showing that releasing Cl creates a vacancy packed with Na ions which is 
a very favorable site for water to adsorb.
Therefore, the release of Cl is only a ``byproduct" of the stronger interaction of water and Na ions.
Finally, we find that there is no significant change in the displacement energy once the displaced ion
and the vacancy are covered with water molecules.

In the next section we discuss the technical details of the methods employed 
to simulate the systems and to obtain representative structures. 
This is followed by results of adsorption on the different surfaces in Section~\ref{sa_sec_flat}. 
The dependence of the energy required to displace the ions is reported in Section~\ref{sa_sec_vac} 
which is followed by discussion and conclusions in Section~\ref{sa_sec_dis}.

\section{Computational procedure and setup}

\begin{figure}[h]
\centerline{
\includegraphics[height=3.9cm]{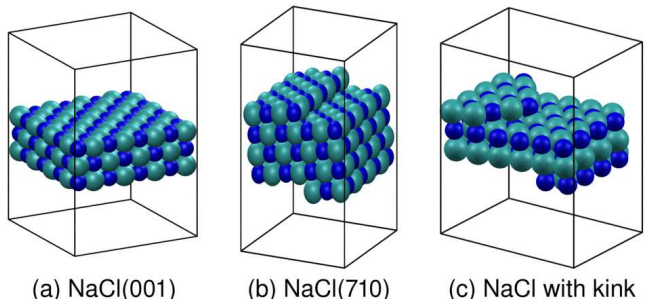}}
\caption[Models of the different NaCl surfaces.]
{Simulation cells used in the study: (a) flat surface, (b) surface with a monoatomic step, and
(c) surface with a kink. Each of the cells shown is repeated in three dimensions using periodic
boundary conditions.
Na, Cl ions are colored blue, green, respectively. 
This color scheme is used throughout.}
\label{fig_sal_box}
\end{figure}

In this study we used a combination of FF based MD as a means to obtain possible structures of the system 
followed by geometry optimization with DFT for water clusters of different size on three different NaCl surface models.
The surfaces were modelled using a slab geometry with periodic boundary conditions. 
We have studied flat NaCl(001), a surface with a monoatomic step (represented by the NaCl(710) surface),
and surfaces with a kink and corner sites in the step (called simply ``surface with a kink" for brevity in the following), 
either Na or Cl terminated. 
The models as well as the simulation cells are depicted in Figure~\ref{fig_sal_box}.
In the NaCl(001) case we used three atomic layers of NaCl and a large $4\times4$ supercell. 
Four atomic layers were used to model NaCl(710), with the layers tilted to create a stoichiometric 001-like step
on the surface (one on each side of the slab); this particular type of structure has been identified as the lowest 
energy step defect.\cite{li2007}
A similar strategy was employed to create the kinks, which leads to one side of
the surface having a Na terminated kink while the other side has a Cl terminated kink. 
Tests performed in this and previous work show that these thin slabs yield sufficiently converged
adsorption and displacement energies.\cite{li2007,li2008,klimes2010jpcm}

The adsorption energy $E_{\rm ads}$ is calculated using
\begin{equation}
E_{\rm ads}=(E^{\rm tot}_{\rm nH_2O/NaCl}-nE^{\rm tot}_{\rm wat}-E^{\rm tot}_{\rm surf} )/ n\,,
\label{eq_salt_Eads}
\end{equation}
where
$E^{\rm tot}_{\rm nH_2O/NaCl}$ is the total energy of the system with a water cluster adsorbed on the surface,
$E^{\rm tot}_{\rm wat}$ is a reference total energy of an isolated water molecule in the gas phase,
$E^{\rm tot}_{\rm surf}$ is the total energy of a water-free relaxed surface.
 $n$ is the number of adsorbed water molecules.

The displacement energy $E_{\rm dis}$ is calculated as
\begin{equation}
E_{\rm dis}=E^{\rm tot}_{\rm nH_2O/NaCl-dis}-E^{\rm tot}_{\rm nH_2O/NaCl}\,,
\label{eq_salt_Evac}
\end{equation}
where $E^{\rm tot}_{\rm nH_2O/NaCl-dis}$ is the total energy of the system with the water cluster adsorbed on the surface with
displaced ions. 
The $E^{\rm tot}_{\rm nH_2O/NaCl}$ is the total energy of the lowest energy adsorbed water
cluster on the surface for that particular number of water molecules.
Therefore the $E_{\rm dis}$ gives the energy cost to move an ion from the surface layer to a less coordinated site. 
A positive number means that displacing the ion is not favorable, whilst a negative $E_{\rm dis}$ represents a release of energy  
and a favorable dissolution process.
Note that here we are dealing with the energetics of this process at zero Kelvin and that thermal and entropic effects
are not taken into account.

For the DFT calculations VASP 4.6.34\cite{kresse1993,kresse1996} was used with the PBE
exchange-correlation functional.\cite{perdew1996}
A plane wave basis set cut-off was set to 400~eV and the large cells employed and the electronic structure 
of the system allows us to use $\Gamma$ point only $k$-space sampling.
The core electrons in VASP are treated within the projector augmented-wave (PAW) method.\cite{blochl1994, kresse1999}
The geometry optimizations were run with maximal atomic force criterion of 0.02~eV/\AA.
While PBE has been shown to describe the binding between water molecules well,\cite{santra2007,santra2008}
it has been shown that it underestimates the adsorption energy of a water monomer on NaCl(001).\cite{li2008} 
This can be, at least partially, attributed to the missing description of dispersion within PBE which can be 
an important part of the adsorption energy.\cite{klimes2010}
Therefore, for some of the structures, we compared the PBE results to results obtained with the optB86b-vdW 
functional,\cite{klimes2010,klimes2011} which includes a long-range correlation part.\cite{dion2004}
While the absolute adsorption energies become larger, the relative differences, mostly governed by stronger
electrostatic contributions, do not change so as not to alter the conclusions drawn with the PBE functional, 
see SI.\cite{supplementary}
This is consistent with recent studies of adsorption of water and other small molecules or clusters 
on surfaces.\cite{carrasco2011Me,carrasco2013,hanke2012,forster2012}

The FF calculations were performed using the Gromacs suite of programs.\cite{lindahl2001, hess2008} 
The AMBER parameters\cite{wang2000,sorin2005} 
were used for Na and Cl and water was modelled using the TIP4P\cite{jorgensen1985} potential
with some tests performed using the TIP4P/2005 parametrization.\cite{abascal2005}
The AMBER and TIP4P parameters were selected since they give the best agreement with PBE data for
various properties of interest from the force fields available in Gromacs, see SI.\cite{supplementary}
For example, the adsorption energies on salt obtained with AMBER and TIP4P are similar to those obtained
from PBE 
(see Figure~S1).
We note that FF calculations were performed simply to generate appropriate structures for subsequent 
PBE calculations and that all energies reported in this article have been obtained with PBE. 
We have chosen PBE over a purely FF-based approach mainly because in our preliminary tests 
we did not identify a combination of FFs (either polarizable or non-polarizable) 
that yielded accurate structures, displacement energies, 
and relative energies for small water clusters when compared with PBE, see SI.\cite{supplementary}

Our approach in this work has been to compare the energy of a cluster adsorbed on the intact surface with the energy
of a cluster on a surface where one or two of the ions have been displaced.
For this we need to have
a sufficient (\ie, large enough) sample of representative low energy structures for each case.
There are several approaches suitable for producing a large number 
of distinct clusters such as simulated annealing, various Monte Carlo based approaches,\cite{wales2003} random
structure searching,\cite{pickard2011} or other techniques. 
We opted for a relatively simple and mostly automated procedure 
where for a given number of water molecules and type of defect we perform 2~ns of FF-MD simulations at 
200, 230, 260, and 300~K. From each of these runs we select a large number ($>$200) of structures and 
perform geometry optimization using Gromacs. By ordering the structures according to their
energy we obtain low energy structures for each combination of defect and number of water molecules, and for some of 
the structures we then perform geometry optimizations using VASP. 
This gives us several PBE optimized structures from which we select the lowest energy one to be used in the
displacement energy calculation.
Our selection procedure is able to give us a representative set of low energy structures. 
However, when the number of water molecules is large, we might not necessarily find the lowest energy PBE cluster 
and free energy calculations should be more appropriate to describe the energy differences.
Although steps in this direction have been performed in our group using DFT,\cite{liu2011} 
a large scale application as presented here\cite{foot_scale} is not yet possible with DFT.
In a separate study on a similar system using FFs we have compared the displacement free energies with the energy
differences of the lowest energy clusters and found rather good agreement.\cite{cheryl_msc2010}

\section{Results}

\subsection{Water adsorption on surfaces without displaced ions}
\label{sa_sec_flat}

\subsubsection{Flat surface}

\begin{figure}[ht]
\centerline{
\includegraphics[height=6.1cm]{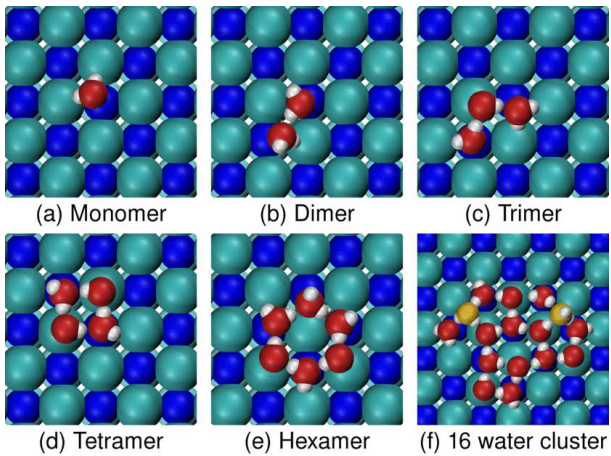}}
\caption[Water clusters on the NaCl(001).]
{The lowest energy water clusters found on a flat surface containing 1, 2, 3, 4, 6, and 16 water molecules.
The water molecules tend
to bind to Na ions but adsorption on Cl is possible because such water molecules can accept hydrogen bonds.
This leads to the formation of loops containing four or more molecules. For large clusters, some of the molecules
can occupy positions in a second layer and are not in direct contact with the surface.
These are shown in orange in (f).
}
\label{fig_sal_f16}
\end{figure}

\begin{figure}[ht]
\centerline{
\includegraphics[height=9cm, angle=-90]{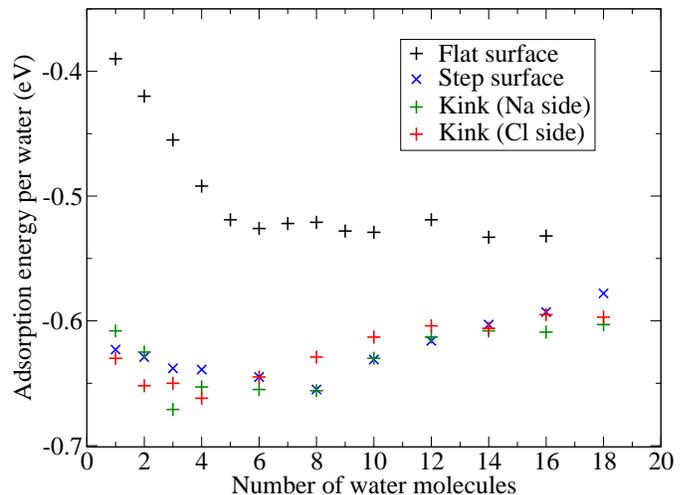}}
\caption[Adsorption energy of water clusters on different NaCl surfaces calculated using DFT-PBE.]
{Adsorption energy per water molecule in eV for the four surfaces studied: Flat (001), monoatomic step (710),
and two surfaces with a Na or Cl kink.}
\label{gra_sal_free2}
\end{figure}

First we study how water molecules adsorb on the selected surfaces trying to understand the general features
of the water substrate interactions.
On the flat surface (Figure~\ref{fig_sal_f16}) the lowest energy structure of a water monomer ($E_{\rm ads}=-390$~meV) 
has the molecular plane almost parallel to the surface with the O atom on top of a Na ion with 
the water hydrogens directed towards adjacent Cl ions,\cite{park2004,cabrera2006,yang2006,li2008,klimes2010jpcm}
see Figure~\ref{fig_sal_f16}(a). 
The water molecule can also bind to the Cl ion 
via a hydrogen bond, although less strongly\cite{park2004,yang2006} ($E_{\rm ads}=-160$~meV).
These two structures can form a dimer where additional energy is obtained through the formation 
of a hydrogen bond.
Adding more waters leads to a further increase in the adsorption energy up to the hexamer, beyond which the adsorption 
energy levels off at around $-570$~meV (Figure~\ref{gra_sal_free2}).
Small clusters (up to tetramer) have been studied previously\cite{park2004,yang2006,cabrera2006,cabrera2007}
and for these we find reasonable agreement with the published adsorption energies and structures.

The leveling off of $E_{\rm ads}$ at around six water molecules is interesting and worth explaining: 
The magnitude of  $E_{\rm ads}$ is determined  by the interactions of the water molecules between each other
and with the surface, a single water molecule can usually form up to four such bonds.
When the clusters are small, not all the possible bonds are created as there is 
an insufficient number of water molecules to bind to. 
For example, in the dimer structure (Figure~\ref{fig_sal_f16}(b)) there are two hydrogens, one on each molecule, 
which do not form any bond.
However, as the cluster size increases, the average number of bonds also increases and at some point 
most of the water molecules participate in four bonds leading to the saturation of $E_{\rm ads}$.

\subsubsection{Surface with a step}

\begin{figure}[ht]
\centerline{
\includegraphics[height=6.1cm]{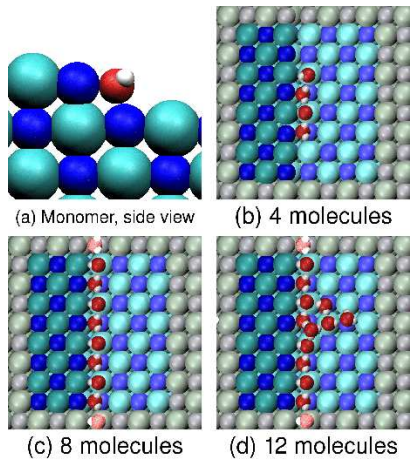}}
\caption[Small water clusters adsorbed on NaCl surface with a monoatomic step.]
{Side view of the water monomer and a top view of water clusters with four, 
eight, and twelve molecules on the surface with a monoatomic step.
Apart from the monomer image the upper terrace is shown with darker colors, the lower with pale colors to highlight the step.
The periodic cells are shown with greyed out colors and this coloring scheme is used throughout.}
\label{fig_sal_step_free}
\end{figure}

Let us now focus on the defective (non-flat) surfaces where a step or a kink is present. 
Although such surface features will inevitably be present in a real crystal, water adsorption on such surfaces 
has not been widely studied 
in the case of NaCl.
In agreement with previous work,\cite{ahlswede1999,boli2009} we find that on the step surface the 
most favorable adsorption site is next to the step where two Na ions are exposed:
one Na ion is below the water molecule and a second is in the step, see Figure~\ref{fig_sal_step_free}(a).
The adsorbed water molecule can bind to both ions which results
in a much more favorable adsorption energy ($E_{\rm ads}=-623$~meV) when compared to the flat surface. 
When more water molecules are added, chain structures are formed, the adsorption energy, however, is 
only marginally affected (e.g., the $E_{\rm ads}$ of a dimer is $-629$~meV).
In this case there is only a small energy difference between water binding to Na
or binding to Cl and accepting a hydrogen bond from another water bound to Na.
For eight water molecules a water chain along the step is formed.
This structure, shown in Figure~\ref{fig_sal_step_free}(c), has the most favorable adsorption energy per water molecule
from all the models on the step surface ($E_{\rm ads}=-655$~meV).
In our calculations the particular number of water molecules adsorbed at the step is given
by the periodic boundary conditions we employ wherein our model
of the step has eight adsorption sites per unit cell.
The structure then represents an infinite water chain adsorbed along the step.
Additional water molecules have to adsorb on the less favorable terrace sites and they need to distort
the chain structure which reduces the adsorption energy.
It is worth noting that similar chain structures have been observed in the experiments\cite{dai1997,luna1998,ghosal2005,verdaguer2005}
and predicted to exist for water on steps of metal surfaces.\cite{morgenstern1996,meng2004,donadio2012} 
However, the structure on the NaCl step differs as it contains two different geometries of water molecules, one type
bonded to Na and the other to Cl.

\subsubsection{Surfaces with a kink}
\begin{figure}[ht]
\centerline{
\includegraphics[height=4.0cm]{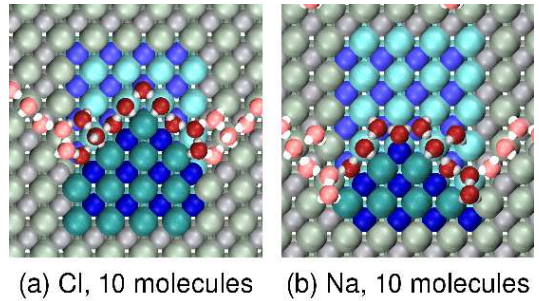}}
\caption{Water clusters with ten molecules adsorbed on NaCl surfaces with a 
Cl terminated kink site (a) and a Na terminated kink site (b).
Similar to the surface with a step, these surfaces exhibit favorable adsorption sites
involving two adjacent Na ions. Moreover, the surface can also support the formation of a water chain.}
\label{fig_sal_kiCl}
\end{figure}

The Na or Cl terminated kink surfaces also contain a step and as one can see in Figure~\ref{gra_sal_free2} 
the adsorption energies are similar to those observed on the step surface.
In the case of the Cl terminated kink, the most stable configuration for a single water is at the Na ions
of the step with $E_{\rm ads}=-630$~meV. 
In a similar manner to the step surface, when more water molecules are added, chain structures are usually the most
stable structural motifs.
Interestingly, the corner site with three Na ions does not facilitate a favorable connection of two chains 
(see Figure~\ref{fig_sal_kiCl}(a) and S7) and the adsorption energy for intermediate number of water molecules (between 6 and 14)
is smaller than on the step surface.
For even larger number of water molecules, the terrace sites are occupied.

On the Na terminated kink (Figure~\ref{fig_sal_kiCl}(b) and S8) the corner site represents a very favorable adsorption site,
for example, three molecules can adsorb with $E_{\rm ads}=-671$~meV.
On the other hand the lowest energy configuration for a monomer is at the terminal Na ion, but with a less favorable $E_{\rm ads}$
of $-608$~meV.
When the number of water molecules is increased, chains are again formed with a favorable structure for eight water molecules.
Between 8 and 14 water molecules, the adsorption energy follows closely what we found on the stepped surface (Figure~\ref{gra_sal_free2}). 
The adsorption is more favorable on the Na terminated kink for 16 and 18 water molecules than on the step 
since on the kink the clusters occupy the corner area between the two periodic kinks and can form more compact structures.

\subsection{Moving ions out of the lattice}
\label{sa_sec_vac}

\subsubsection{Flat surface}

We now explore the displacement of ions from the lattice, starting with the flat surface. 
Although this surface is not expected to be the easiest place to initiate dissolution, 
it turns out that we can learn a lot about the general factors relevant to the process.
Let us just repeat that we study how the displacement energy $E_{\rm dis}$ for different defects depends on the 
number of water molecules adsorbed.
This gives us the energetic cost to displace the ion and form the vacancy and how this is affected by the presence of water. 
Only when $E_{\rm dis}$ is negative is it favorable to release the ion from the lattice and form the vacancy.

\begin{figure}[ht]
\centerline{
\includegraphics[height=6.1cm]{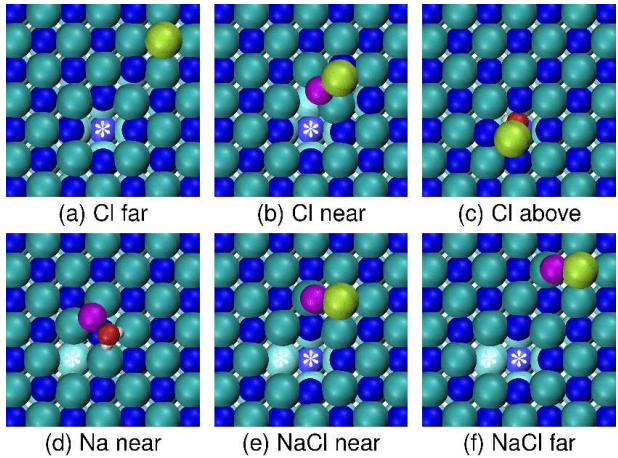}}
\caption[Displacement defects on flat NaCl(001).]
{Displacement defects on flat NaCl(001): 
(a) ``Cl far" with the Cl ion far from the vacancy, 
(b) ``Cl near" where the displaced Cl ion occupies a site near the vacancy and can displace an adjacent undercoordinated Na ion, 
(c) ``Cl above" where one water molecule prevents the displaced Cl from returning to its original lattice site, 
(d) ``Na near" where the Na ion has been taken out, 
(e) ``NaCl near", an ion pair defect proximate to the vacancy and (f) ``NaCl far" with the ions taken farther.
For clarity vacancies are generally indicated with an asterisk.}
\label{fig_sal_fla_vac}
\end{figure}

We compare the energetic cost of creating various types of defects: single Cl or single Na ions or a NaCl ion pair, 
shown in Figure~\ref{fig_sal_fla_vac}.
For the flat surface we create different Cl defects by placing the ion two lattice positions 
(``Cl far", Figure~\ref{fig_sal_fla_vac}(a)) or one lattice position away (``Cl near") from its original site.
Often when the Cl near structure is created, an adjacent Na ion is pulled out from the surface, 
as shown in Figure~\ref{fig_sal_fla_vac}(b). 
The last Cl defect ``Cl above" (Figure~\ref{fig_sal_fla_vac}(c)) is formed when the Cl ion
occupies the position above the vacancy but the vacancy itself is occupied by a water molecule.
The Na ion can form a ``Na near" defect, shown in Figure~\ref{fig_sal_fla_vac}(d).
It doesn't form a structure corresponding to the ``Cl above" for reasons
we will discuss later and the ``Na far" defect becomes stable in the FF-MD simulations only for a large number of water molecules.
The ion pair can be removed and displaced to a position closer or farther from the vacancy leading to 
``NaCl near" and ``NaCl far" structures, shown in Figures~\ref{fig_sal_fla_vac}(e) and (f), respectively.

\begin{figure}[ht]
\centerline{
\includegraphics[height=9cm, angle=-90]{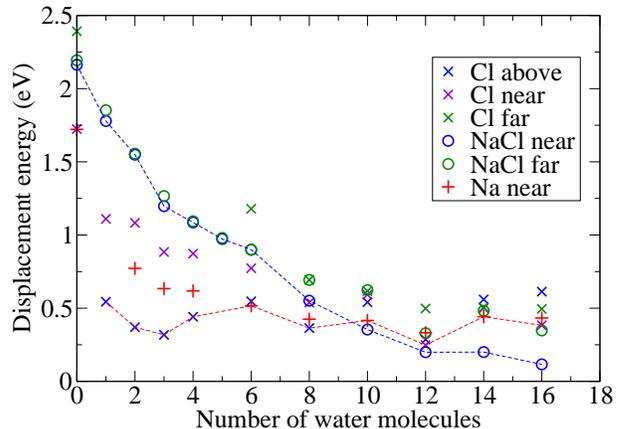}}
\caption[Defect formation energies on the flat surface.]
{The displacement energy in eV for different defects and number of water molecules on the flat surface.
The value represents the energy difference between the lowest energy water cluster adsorbed on a flat surface
and the lowest energy structure identified for a cluster with the same number of water molecules plus a defect.
The red (blue) dashed curves connect the lowest energy single ion (ion pair) defects.
It can be seen that for ten water molecules and more ion pair defects are favored over single ion defects.}
\label{gra_sal_vflat}
\end{figure}

In the absence of water the displacement energies are all quite high, e.g., 1.7~eV for a single 
Cl defect and 2.2~eV for the ion pair defect (Figure~\ref{gra_sal_vflat}).
However, as can be seen in Figure~\ref{gra_sal_vflat}, in the presence of water molecules it becomes
significantly easier to create defects.
This is simply caused by the fact that the surface with the defect offers more favorable adsorption sites 
for water molecules, \ie, sites where the absolute value of $E_{\rm ads}$ is larger than on the flat surface.
Such favorable sites occur not only in and around the vacancy but also around the displaced ion.
In the case of a Cl ion defect, the water molecule adsorbs directly into the vacancy where many Na ions are exposed.
This leads to an interesting situation as the water can prevent the ion from coming back into the vacancy, 
forming the ``Cl above" structure with a relatively small $E_{\rm dis}$.
In the case of the Na defect the water adsorbs preferentially at the displaced ion. 
Clearly in both cases the behavior of water is governed by the strong water-Na interaction.
However, as there is no water directly in the vacancy the Na ion can return.
From the energetics established of the single ion defects it's clear that there is a competition between the ion-ion 
and ion-water interactions.
The water molecules interact strongly with Na ions and preferentially adsorb at them;
the Na ions also are less ``stable" when removed from the lattice site and can return to the vacancy.

\begin{figure}[ht]
\centerline{
\includegraphics[height=3.2cm]{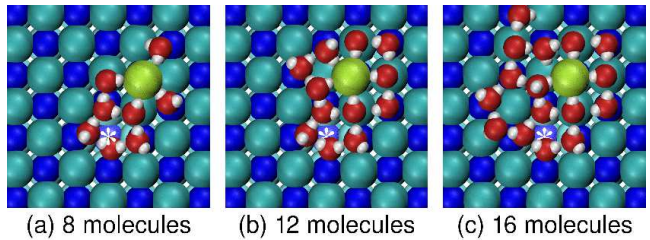}}
\caption[The ``Cl near" defect with different number of water molecules.]
{The ``Cl near" defect on the flat surface with 8, 12, and 16 water molecules.}
\label{fig_sal_fla_12}
\end{figure}

When we look at $E_{\rm dis}$ for more than 8 water molecules we note that 
the energy cost needed to create the single ion defect basically does not change. 
This can again simply be understood from Figure~\ref{fig_sal_fla_12}, where a ``Cl near" defect
with 8, 12, and 16 waters is shown. 
Once the vacancy and displaced ion are covered by water molecules the defects are ``healed"
and subsequent water molecules to absorb see an environment similar to adsorption on a flat surface without any defect.
Therefore $E_{\rm dis}$ does not change appreciably when additional water molecules are adsorbed. 
This is an important point, the displaced ion or vacancy enhance
the adsorption energy only for the water molecules directly in contact with them.
While there is little change in $E_{\rm dis}$ for the single ion defects for more than 8 water molecules, 
the cost to create the pair defects is still reducing.
For 10 water molecules the pair defects become less costly than the single ion defects and
for the largest clusters we have studied  $E_{\rm dis}$ drops to just 100--200~meV.
Again, the cost of pair defects is initially higher since more bonds need to be broken,
but since the structures contain more favorable adsorption sites than the single ion defects
the final $E_{\rm dis}$ becomes smaller.
However, as the number of water molecules is increased above 12, there is little change in  $E_{\rm dis}$
as there are only few favorable adsorption sites left.
Thus on the flat surface we have not observed a situation where it would be energetically favorable 
for the ion to leave the lattice. 

\subsubsection{Surface with a step}

\begin{figure}[ht]
\centerline{
\includegraphics[height=6.1cm]{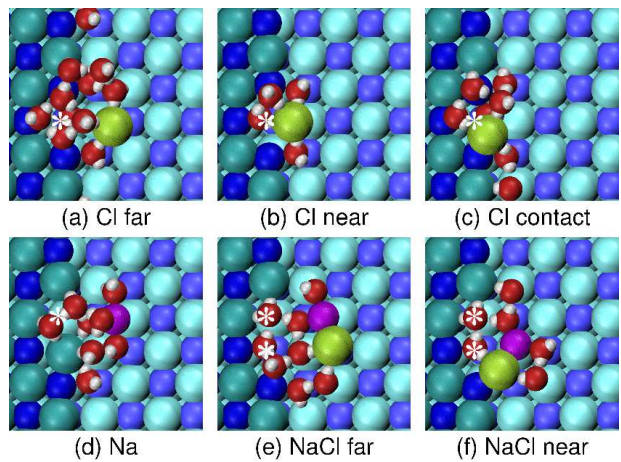}}
\caption[Defects on the NaCl surface with a monoatomic step.]
{Defects on the NaCl surface with a monoatomic step. The Cl ion can reside far (``Cl far")
from the vacancy, or can sit in a hydration shell (``Cl near"), or it can come into contact with a Na ion adjacent 
to the vacancy site (``Cl contact"). The Na ion defects are less stable and we only show the structure
for the ``Na" defect,
which is equivalent to the ``Cl far". We displace the ion pair to a ``NaCl far" position, and also structures denoted
as ``NaCl near" occurred during the molecular dynamics, in these structures the Cl ion is raised above the step.}
\label{fig_sal_st_vac}
\end{figure}

We now focus on the surface with a monoatomic step, where some of the ions are less coordinated. 
As with the flat surface we have compared the energy cost to create single Na or Cl defects, 
and ion pair defects displaced by various amounts from their original lattice positions, see Figure~\ref{fig_sal_st_vac}. 
On the stepped surface we identified a defect that we call ``Cl contact", shown in Figure~\ref{fig_sal_st_vac}(c), which is similar to the
``Cl above" on the flat surface, i.e., the Cl ion is close to the vacancy but hindered from returning to it by a water molecule.

\begin{figure}[ht]
\centerline{
\includegraphics[height=9cm, angle=-90]{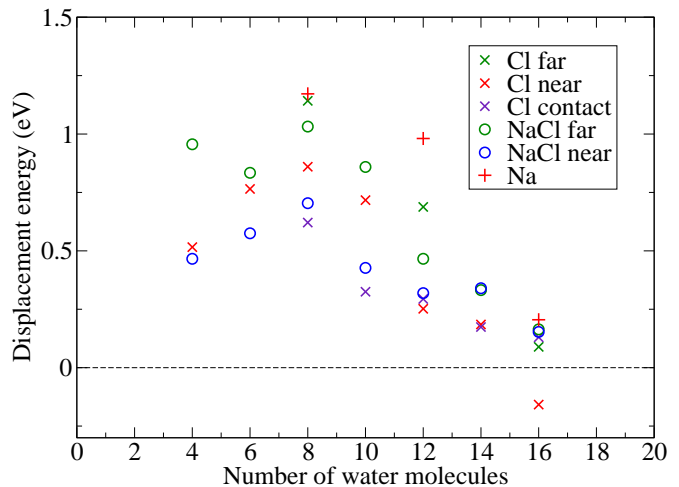}}
\caption[The defect formation energy on a surface with a monoatomic step.]
{The displacement energy on a surface with a monoatomic step. We show data for a single Cl or Na defect
in different arrangements and data for two types of NaCl dimer defects. As can be seen for 16 water molecules
it is more favorable to form the ``Cl near" defect than to form a water cluster on a defect-free step.}
\label{gra_sal_vstep}
\end{figure}

\begin{figure}[ht]
\centerline{
\includegraphics[height=3.2cm]{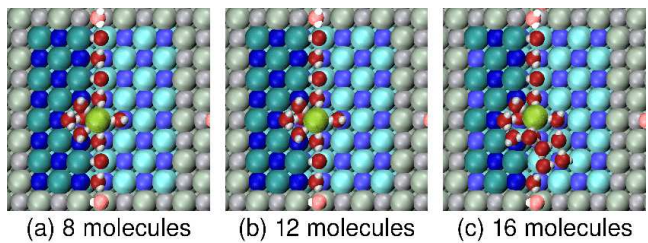}
}
\caption[Monoatomic step with a ``Cl near" defect.]
{The hydrated Cl ion defect (``Cl near") on the monoatomic step with eight, twelve, and sixteen water molecules.
This defect represents a very stable structure and does, in fact, become more stable than the 
perfect step for 16 water molecules, as indicated in Figure~\ref{gra_sal_vstep}.}
\label{fig_sal_stF4}
\end{figure}

As can be seen from Figure~\ref{gra_sal_vstep} $E_{\rm dis}$ first rises when water molecules are added. 
Although this at first might seem surprising it is simply caused by the strong adsorption of water on the step.
For more than eight water molecules the step without any defect does not represent such a favorable adsorption site 
and $E_{\rm dis}$ quickly decreases when water molecules are added.
One particularly stable structure is formed by the ``Cl near" defect where the Cl ion is taken out 
of the step and partially hydrated by water molecules that bind to the exposed Na ions at the same time
(see Figure~\ref{fig_sal_stF4}). 
The energy gain obtained from the strong adsorption on the Na ions is large enough to bring $E_{\rm dis}$
below zero for 16 water molecules, \ie, it is energetically favorable to release the Cl ion.
For all the systems examined in this study this is the only configuration identified where it becomes
favorable to remove an ion from its lattice site.
This is in agreement with the SPFM measurements which suggest that the Cl ions leave the step defects 
first.\cite{garcia2004,ghosal2005,verdaguer2005,verdaguer2008}

The ion pair defects are less costly to form than on the flat surface with the cost being similar to that of the single Cl
defects. 
It seems plausible 
from the trend shown in Figure~\ref{gra_sal_vstep} that the displaced NaCl pair 
will become favorable for more water molecules than we have examined here ($>$16).
The Na defect by itself is not favorable to form, even with respect to the ion pair defects,
which agrees with our former findings.
Indeed, when Na is displaced from the step, 
a site packed with Cl ions is exposed which does not yield favorable adsorption sites for water. 

\subsubsection{Surfaces with a kink}

\begin{figure*}[ht]
\centerline{
\includegraphics[height=6.5cm, angle=-90]{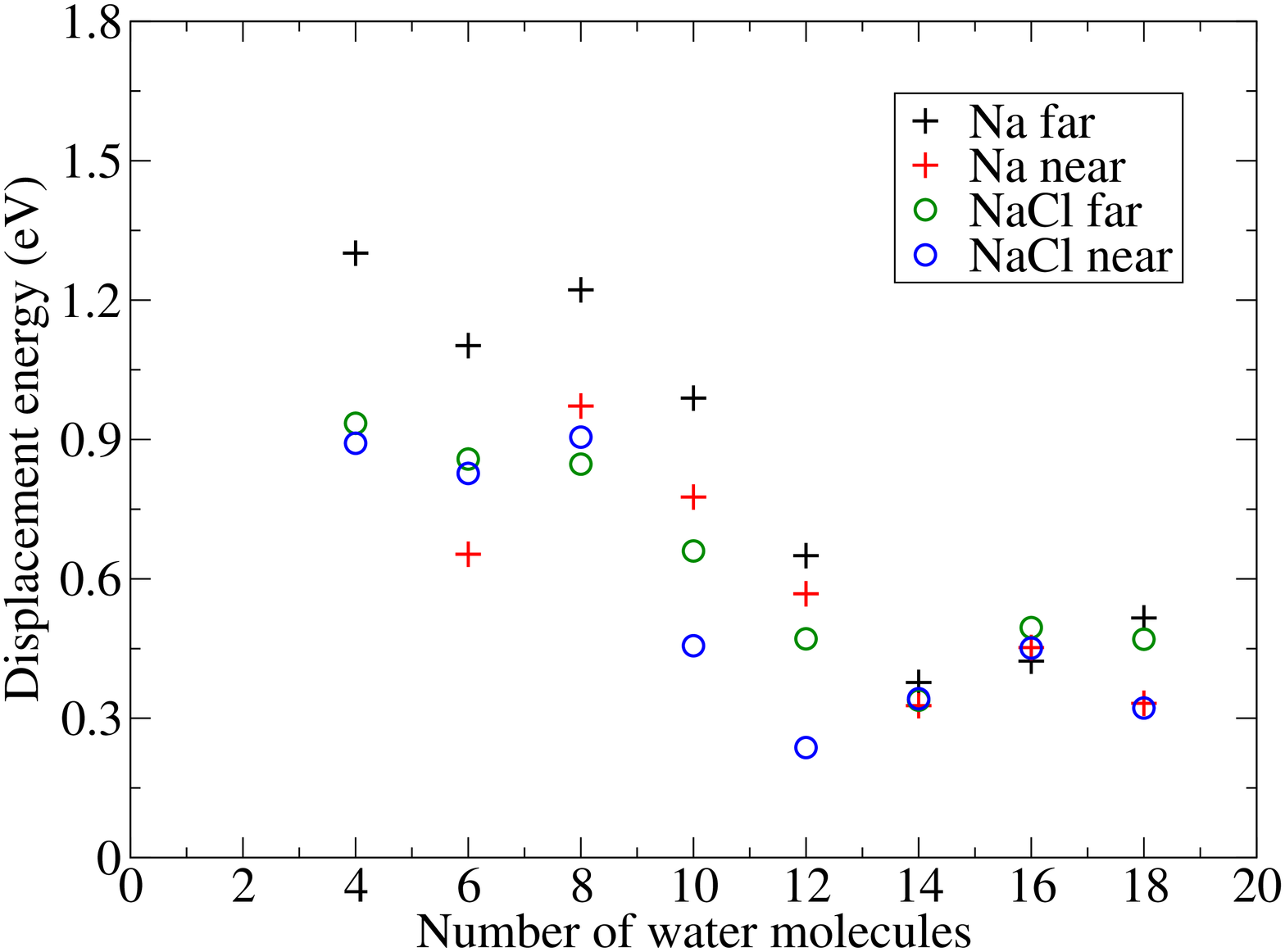}
\includegraphics[height=6.5cm, angle=-90]{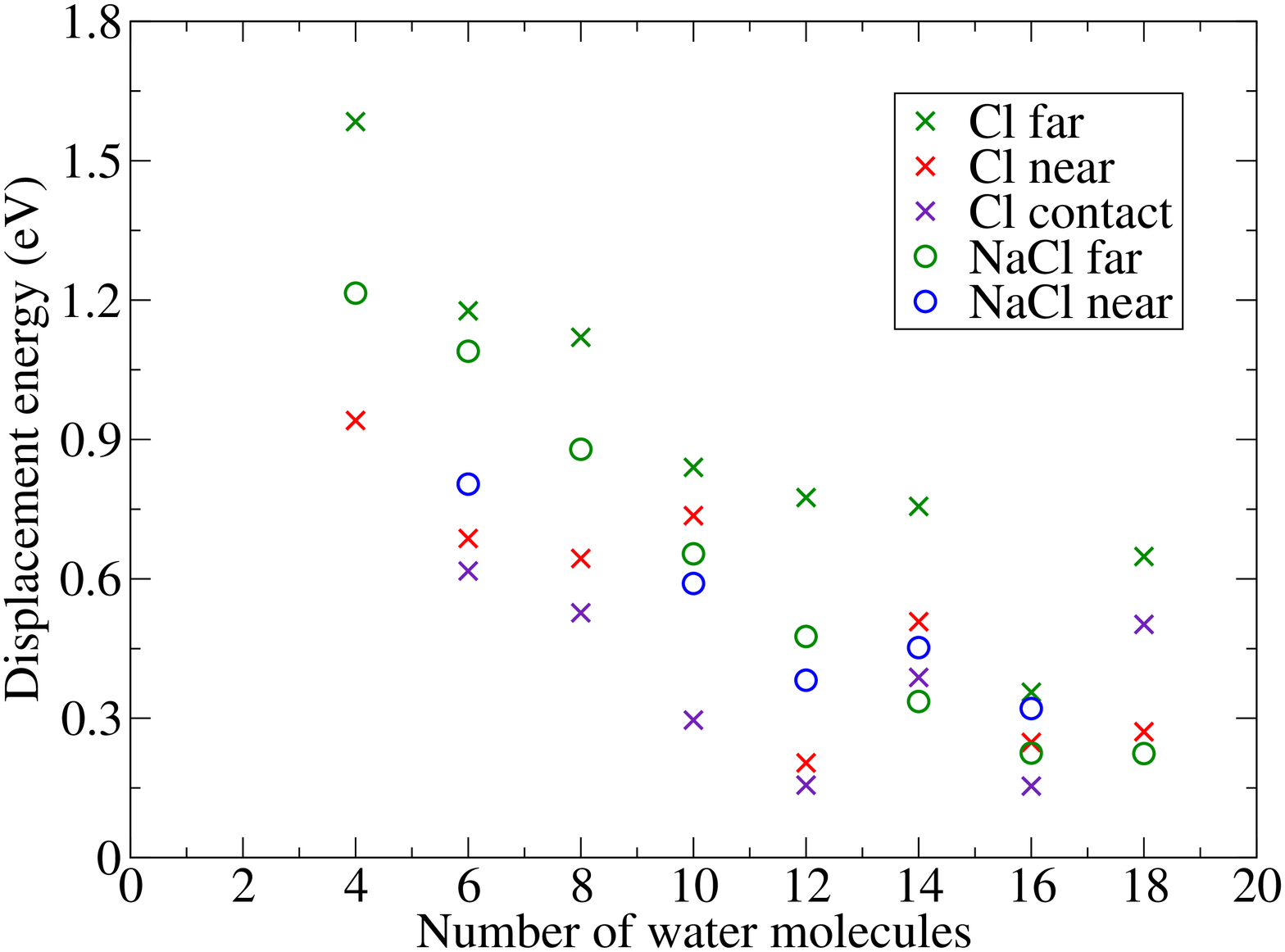}}
\caption[Defect formation energies on Na terminated kink.]
{The displacement energy on Na (left) and Cl (right) terminated kink surfaces.}
\label{gra_sal_vkink3}
\end{figure*}

We now consider the surfaces with either Na or Cl terminated kink sites. As with the flat and stepped surfaces 
we calculate $E_{\rm dis}$  for a single ion or ion pair defects and various numbers of water molecules. 
In the case of a Na terminated kink
the single ion defect is a Na ion, which upon displacement exposes two Cl ions, shown in Figure~S9(a).
As we have already learnt from the previous surfaces such a structure is not favorable for water adsorption 
and one would not expect the Na defects to be preferred. 
Figure~\ref{gra_sal_vkink3} shows that this is indeed the case and in fact it is easier to create 
the ion pair defects at this particular surface. 
However, even for the pair defects $E_{\rm dis}$ for a large number of water molecules is quite high, above 300~meV.
The increase in $E_{\rm dis}$ seen in Figure~\ref{gra_sal_vkink3} for 14--18 water molecules 
is partially caused by a lower adsorption energy on the kink visible in Figure~\ref{gra_sal_free2}, another
reason could be insufficient sampling of the lowest energy clusters.

The last surface model we have considered is the Cl terminated kink structure. The different
defects considered are shown in Figure~S10 and the energies shown in Figure~\ref{gra_sal_vkink3}.  
The displacement of Cl ions exposes favorable adsorption sites for water molecules and thus 
one would expect that: i) the release of the ions could become favorable for a large number of waters;
and ii) the cost of single ion defects will be initially lower than the cost of ion pair defects
(as opposed to the situation on the Na terminated kink which we just discussed). 
The latter point is indeed observed:
the single Cl ion defect in the ``Cl near" configuration
has a lower $E_{\rm dis}$ than the ``NaCl near" defect (see Figure~\ref{gra_sal_vkink3}). 
Moreover, the structure ``Cl contact", where the Cl ion is only slightly displaced 
has an even lower energy cost. 
However, it seems that this defect does not expose enough Na ions to yield a favorable 
solvation structure compared to the kink surface without a defect.
Although the final $E_{\rm dis}$ for most of the defects are quite low, around 200~meV, 
we haven't found a structure that is more stable than the defect free kink surface.\cite{foot_kink_clus}

\section{Discussion and Conclusions}
\label{sa_sec_dis}

We have performed a detailed study aimed at elucidating the intimate details
of the initial stages of salt dissolution in water.
The different surface models used enabled us to find basic principles which govern the
process and we discuss our findings in more detail here.

Let us start with the behavior of water on surfaces without displaced ions.
As observed previously water molecules bind more strongly to the Na ion than to the Cl ion.
Furthermore the binding is even stronger when two adjacent Na ions are exposed, for example on the step.
Adsorption at Cl sites is more likely when the adjacent Na ions are occupied in which case 
the two molecules form a hydrogen bond with each other.
This leads to formation of clusters on the flat surface and molecular chains that run along the steps
for the defective surfaces.
As can be seen from Figure~\ref{gra_sal_free2} chains of water molecules at step sites are much more stable than clusters
on the flat surface.
However, once water covers the favorable sites on the step, further water molecules need to 
adsorb at the less favorable sites at the terrace and the average adsorption energy is reduced.
These findings are in agreement with experimental observations in which water was found
to adsorb first on defects, such as steps.\cite{dai1997,luna1998,ghosal2005,verdaguer2005}

\begin{figure}[ht]
\centerline{
\includegraphics[height=5.0cm]{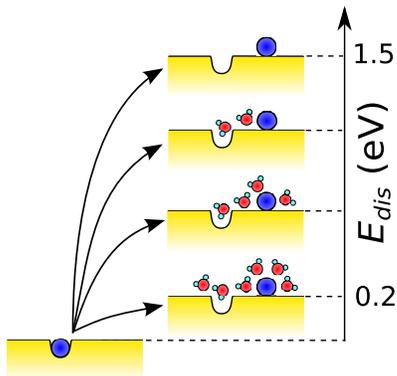}
}
\caption[Schema of the energy trends.]
{Schematic illustration of the main trends in the displacement energy $E_{\rm dis}$.
The energy cost to create the defect is high but it also creates favorable adsorption sites for water
both at the vacancy and at the displaced ion.
As more energy is gained by adsorbing at the defect, the cost of creating the defect is reduced.
}
\label{fig_schema}
\end{figure}

The energetic cost required to release an ion from the lattice is governed by two competing forces:
i) the energy lost in creating a defect (breaking ionic bonds); and 
ii) the energy gained by adsorbing water at the vacancy and around the displaced ions.
This is schematically illustrated in Figure~\ref{fig_schema}.
The cost to create a defect is high, about 1--2~eV both for a single ion or a pair, 
but the adsorption energy of water is enhanced by $\sim$0.2~eV for each water that adsorbs at a defect.
(This is approximately the difference between $E_{\rm ads}$ on the flat surface and on the surfaces 
with defects, see Figure~\ref{gra_sal_free2}.)
Thus the release of ions from the crystal is not favorable but adding water molecules reduces the cost through the stronger binding to the defect.
However, once the defects are covered by water $E_{\rm dis}$
no longer decreases when more water molecules are added since all the favorable adsorption sites are occupied.
This can be clearly seen for the flat surface single ion $E_{\rm dis}$ in Figure~\ref{gra_sal_vflat}.
For cases where there are more water molecules than defect sites to bind to additional ions can be released.
This is seen on the flat surface where for more than 10 water molecules ion pair defects are preferred.

An interesting result to come out of this study is the role the Na ions play in dictating  
which ion will leave the lattice first.
We note that this is something that the MD studies haven't identified.
As we have seen the water molecules bind more strongly to the Na ions than to the Cl ions.
However, as Fig. 10 clearly shows for the stepped surface, more energy is gained 
when a single Cl is released than when a single Na ion is released.
This is because upon release of the Cl several Na ion adsorption sites around the vacancy are exposed.
Although the energy difference is the main driving force, the Na ions 
also seem to be less ``stable" when removed from their lattice sites.
Taken together this leads to the conclusion that the Cl ions are released first for two key reasons:
i) when the Cl ion is released, favorable Na adsorption sites are exposed; and 
ii) water molecules bind to the vacancy screening it from the displaced Cl ion and also creating a kinetic barrier 
for the Cl ion to return.
Thus the first dissolution step seems to be unexpectedly governed by a rather inverse logic:
it is not the energy gained from removal and hydration of the Cl ion that drives the release of the ion 
but rather the energy gained by exposing more favorable Na adsorption sites.
These conclusions explain the experimental observations that
if there are ions present on the terraces, it will be the Na ions that are hydrated first.
Whilst on steps, however, the Cl ions will be released first so that water can access the Na ions,
in agreement with the experimental observations.\cite{garcia2004,ghosal2005,verdaguer2005,verdaguer2008}
Our findings also agree with the results of the {\it ab initio} molecular dynamics simulations presented in Ref.~\onlinecite{liu2011}.
There it was found that the energy required to displace a Cl ion from a kink site and solvate it is lower 
than the energy needed for a Na ion.
The larger polarizability of Cl was identified as a major cause of this effect.
Our results suggest that the different structure of the first solvation shell also plays a role:
while the water-Na ion interaction is stronger, the molecules need to be arranged in a quite
structured manner around the small and less polarizable Na ion.
In contrast, there is more flexibility for the water molecules around the larger and more polarizable Cl ion which 
is reflected by the large variability of Cl defects we observe.
These differences in the solvation shell are likely to affect the entropic contribution to the barrier for dissolution
in to a liquid overlayer, however, the main driving force should, nevertheless, originate from the stronger
interaction with Na ions.
The mostly electrostatic arguments of the driving forces are supported by the fact that changing the DFT
functional from PBE to optB86b-vdW leads to insignificant changes in $E_{\rm dis}$ (see SI).\cite{supplementary}

Let us briefly discuss the question of where the dissolution will start: at the kink or the step?
When we compare the energies to extract a Cl from a step and a kink we see that they
are quite similar but the values at the step are slightly smaller. 
We do not observe the release to be favorable on the kink, despite the fact that the Cl ion is less coordinated.
However, rounding of kinks is observed experimentally for surfaces exposed to humidity or for surfaces
cleaved at humidities around 40~\%.\cite{dai1997,luna1998,garcia2004,ghosal2005,verdaguer2005}
Therefore, a fuller investigation of this issue is required, preferably with an even larger
simulation cell than we have used here, before a definitive answer can be arrived at.
Concerning the question of whether individual ions or ion pairs will be released first, we notice that the ion pair has a very low  
$E_{\rm dis}$ on both the step and kink.
Moreover, the pair does not require a precise structure to be formed as the Cl ion does on the step and it seems plausible 
that the pair becomes favorable with a larger number of water molecules.
However, again for a final answer a larger model for the kink would be required and in such a situation the 
increased number of water molecules would most likely call for 
molecular dynamics simulations to be performed in order to obtain a displacement free energy.
We note that the current models contain already over two hundred atoms and it is beyond the scope  
of the current study to perform the free energy calculations.

There are certainly other interesting questions that emerge from this study, for example, 
how transferable are our findings to the dissolution of other systems, such as other alkali halides? 
Unfortunately, this is not that easy to answer.
In Ref.~\onlinecite{hu2011}, water adsorption on a range of alkali halides was examined and only on LiF and LiCl
was a similar monomer adsorption structure found.
On other substrates, notably those with larger cations, a hydrogen bonded geometry
with the oxygen pointing away from the surface is found (see Figure~1 in Ref.~\onlinecite{hu2011}).
This suggests that different water cluster structures are likely to appear on other alkali halides, possibly leading
to a different ionic release mechanism.
Luna~{\it et al.} studied water adsorption using SPFM not only on NaCl but also on KCl, KBr, and KF.\cite{luna1998} 
They observed some common features, such as preferential adsorption on steps, rounding of kinks upon
exposure to humidity, quick movement of steps, and, eventually, dissolution.
There are, however, differences in the values of relative humidities at which the fast movement and dissolution happen.
Recently, Ghosal~{\it et al.} found that for a Br doped NaCl crystal, Br ions are preferentially released compared 
to the Cl ions.\cite{ghosal2005}
While, based on our observations from this study, one may conjecture that it will be less costly to displace a Br ion
than to displace a Cl ion there might be other effects playing a role.
Studies of water adsorption on stepped surfaces of the other alkali halides would certainly shed more light
on this issue as well as on the behavior of water on the surfaces observed by Luna~{\it et al.}
Furthermore, unlike a previous study\cite{barnett1996} where water dissociation on charged non-stoichiometric clusters
was studied we haven't observed dissociation of water molecules.\cite{foot_disoc}
What one can expect to be a very general property valid for other materials is the observation that
$E_{\rm dis}$ is reduced only up to the point when all the favorable adsorption sites are covered by water.

To conclude we have used a combination of force field molecular dynamics and 
density functional theory to elucidate the details of the initial stages of NaCl dissolution.
We have found that the release of ions is possible at surface defects even with relatively few water molecules.
This agrees with SPFM experiments where similar conclusions have been drawn for relative humidities below $\sim$40~\%.
We find that the process is initially driven by the stronger affinity of water molecules towards Na ions.
This causes the Cl ions to be released first so that more Na ions are exposed, somewhat defying the 
simplistic view and original experimental assumption\cite{luna1998} that Na ions will dissolve first 
because of their stronger interaction with water molecules. 

\begin{acknowledgments}

AM was supported by the EPSRC and the European Research Council.
AM is also supported by the Royal Society
through a Royal Society Wolfson Research Merit Award.
DRB was supported by the Royal Society and JK was supported by UCL and EPSRC through the PhD+ scheme.
We are grateful to the London Centre for Nanotechnology and UCL Research Computing
for computational resources.
Via our membership of the UK's HPC Materials Chemistry Consortium, which is funded by EPSRC (EP/F067496), 
this work made use of the facilities of HECToR, the UK's national high-performance computing service, 
which is provided by UoE HPCx Ltd at the University of Edinburgh, Cray Inc and NAG Ltd, 
and funded by the Office of Science and Technology through EPSRC's High End Computing Programme.
We thank Dr. Erlend R. M. Davidson for help with compilation of VASP on the Cray-X2 ``Black Widow"
system on which a major part of this study was performed.
We thank Dr. Eva Klime\v{s}ov\'{a} for making Figure~\ref{fig_schema}. 
\end{acknowledgments}


\begin{thebibliography}{66}
\bibitem[{foo({\natexlab{b}})}]{foot_kink_clus}
\emph{\bibinfo{title}{{\rm Another reason for not observing favorable release
  of the ion from the kink might be the structural model we have chosen. On the
  kink structure the water molecules can form clusters in the corners between
  the terraces. The adsorption in such clusters seems to be quite strong as can
  be inferred from Figure~\ref{gra_sal_free2}. Between 12 and 18 water
  molecules the adsorption energy on the kink is almost constant while there is
  a progressive loss on the step. Thus it can be expected once the space
  between the steps on the kinks are full of water clusters, the adsorption of
  subsequent water molecules will be less stabilized and the creation of the
  defects more favored. A model of the surface which would not be affected by
  the adsorption around corner sites could lead to change in the results. For
  this, however, either a very large periodic cell would have to be used or a
  cluster model.}}}
\end{thebibliography}

\end{document}